\documentclass{article}
\usepackage{ltwol2e}
\usepackage{epsfig}

\def\gsim{\ \rlap{\raise 3pt \hbox{$>$}}{\lower 3pt \hbox{$\sim$}}\ }
\def\lsim{\ \rlap{\raise 3pt \hbox{$<$}}{\lower 3pt \hbox{$\sim$}}\ }

\begin{document}

\begin{titlepage}

\begin{flushright}
CERN-TH/98-301\\
hep-ph/9809377
\end{flushright}

\vspace{2.5cm}

\begin{center}
\Large\bf 
Theoretical Status of \boldmath$B\to X_s\gamma$\unboldmath\ Decays
\end{center}

\vspace{1.2cm}

\begin{center}
Matthias Neubert\\
{\sl Theory Division, CERN, CH-1211 Geneva 23, Switzerland}
\end{center}

\vspace{1.3cm}

\begin{center}
{\bf Abstract:}\\[0.3cm]
\parbox{11cm}{
We review the theoretical understanding of the branching ratio and 
photon-energy spectrum in $B\to X_s\gamma$ decays at next-to-leading 
order in QCD, including consistently the effects of Fermi motion. 
For the Standard Model, we obtain ${\rm B}(B\to X_s\gamma) 
=(3.29\pm 0.33)\times 10^{-4}$ for the total branching ratio, and  
${\rm B}(B\to X_s\gamma)= (2.85\,\mbox{}^{+0.34}_{-0.40})\times 
10^{-4}$ if a cut $E_\gamma^{\rm lab}>2.1$\,GeV is applied on the 
photon energy, as done in the recent CLEO analysis. A precise 
measurement of the photon spectrum would help reducing the 
theoretical uncertainty and yield important information on the 
momentum distribution of $b$ quarks inside $B$ mesons.}
\end{center}

\vspace{1cm}

\begin{center}
{\sl To appear in the Proceedings of the\\
XXIXth International Conference on High Energy Physics\\
Vancouver, B.C., Canada, 23--29 July 1998}
\end{center}

\vfil
\noindent
CERN-TH/98-301\\
September 1998

\end{titlepage}

\setcounter{page}{1}

\title{THEORETICAL STATUS OF \boldmath$B\to X_s\gamma$\unboldmath\ 
DECAYS}

\author{M. NEUBERT}

\address{Theory Division, CERN, CH-1211 Geneva 23, Switzerland\\
E-mail: Matthias.Neubert@cern.ch}  

\twocolumn[\maketitle\abstracts{
We review the theoretical understanding of the branching ratio and 
photon-energy spectrum in $B\to X_s\gamma$ decays at next-to-leading 
order in QCD, including consistently the effects of Fermi motion. 
For the Standard Model, we obtain ${\rm B}(B\to X_s\gamma) 
=(3.29\pm 0.33)\times 10^{-4}$ for the total branching ratio, and  
${\rm B}(B\to X_s\gamma)= (2.85\,\mbox{}^{+0.34}_{-0.40})\times 
10^{-4}$ if a cut $E_\gamma^{\rm lab}>2.1$\,GeV is applied on the 
photon energy, as done in the recent CLEO analysis. A precise 
measurement of the photon spectrum would help reducing the 
theoretical uncertainty and yield important information on the 
momentum distribution of $b$ quarks inside $B$ mesons.}]

\section{Introduction}

About three years ago, the CLEO Collaboration reported the first
measurement of the inclusive branching ratio for the radiative decays
$B\to X_s\gamma$, yielding~\cite{CLEO} $\mbox{B}(B\to X_s\gamma)
=(2.32\pm 0.57\pm 0.35)\times 10^{-4}$. At this Conference, this 
value has been updated to~\cite{CLEOnew} $\mbox{B}(B\to
X_s\gamma)=(3.15\pm 0.35\pm 0.32\pm  0.26)\times 10^{-4}$, where the
first error is statistical, the  second systematic, and the third
accounts for model dependence. The ALEPH Collaboration has
reported a measurement of the  corresponding branching ratio for
$b$ hadrons produced at the $Z$  resonance, yielding~\cite{ALEPH}
$\mbox{B}(H_b\to X_s\gamma) =(3.11\pm 0.80\pm 0.72)\times 10^{-4}$.
Theoretically, the two numbers are expected to differ by at most a
few percent.  Taking the weighted average gives
\begin{equation}
   \mbox{B}(B\to X_s\gamma)=(3.14\pm 0.48)\times 10^{-4} \,.
\label{BRexp}
\end{equation}

Being rare processes mediated by loop diagrams, radiative decays of
$B$ mesons are potentially sensitive probes of New Physics beyond the
Standard Model, provided a reliable calculation of their branching
ratio can be performed. The theoretical framework for such a
calculation is set by the heavy-quark expansion, which predicts that
to leading order in $1/m_b$ inclusive decay rates agree with the
parton-model rates for the underlying decays of the $b$
quark~\cite{Chay}$^-$\cite{Adam}.  The leading nonperturbative
corrections have been studied in detail and are well understood. The
prediction for the $B\to X_s\gamma$ branching ratio suffers, however,
from large perturbative uncertainties if only leading-order
expressions  for the Wilson coefficients in the effective weak
Hamiltonian are employed~\cite{Gr90}$^-$\cite{Poko}. Therefore, it was
an important achievement when the full next-to-leading order
calculation of the total $B\to X_s\gamma$ branching ratio in the
Standard Model was completed, combining consistently results for the
matching conditions~\cite{Adel}$^-$\cite{Buras}, matrix
elements~\cite{AliGr,Greub}, and anomalous dimensions~\cite{Chet}.
The leading QED and electroweak radiative corrections were included,
too~\cite{CzMa,us}. As a result, the theoretical uncertainty was
reduced to a level of about 10\%, which is slightly less than the
current experimental error.  During the last year, the next-to-leading
order analysis was extended to the cases of  two-Higgs-doublet
models~\cite{Ciuc,Borzu} and supersymmetry~\cite{SUSY}, so that
accurate theoretical predictions are now at hand also for the most
popular extensions of the Standard Model.

The fact that only the high-energy part of the photon spectrum in
$B\to X_s\gamma$ decays is accessible experimentally introduces a
significant additional theoretical uncertainty~\cite{us}. For
instance, in the new CLEO analysis reported at this
Conference~\cite{CLEOnew} a lower cut on the photon energy of 2.1~GeV
in the laboratory is imposed, which eliminates about three quarters of
phase space in this variable.  After reviewing the  theoretical status
of calculations of the total $B\to X_s\gamma$ branching ratio,  we
thus discuss to what extent the effects of a photon-energy cutoff can
be controlled theoretically, pointing out the importance of
measurements of the photon spectrum for reducing the theoretical 
uncertainty in the extraction of the total branching ratio.

\boldmath
\section{$B\to X_s\gamma$ branching ratio}
\unboldmath

The starting point in the analysis of $B\to X_s\gamma$ decays is the 
effective weak Hamiltonian
\begin{equation}
   H_{\rm eff} = -\frac{4 G_F}{\sqrt2}\,V_{ts}^* V_{tb} \sum_i
   C_i(\mu_b) O_i(\mu_b) \,.
\label{Heff}
\end{equation}
The operators relevant to our discussion are
\begin{eqnarray}
   O_2 &=& \bar s_L\gamma_\mu c_L\bar c_L\gamma^\mu b_L \,,
    \nonumber\\ [0.1cm]
   O_7 &=& \frac{e\,m_b}{16\pi^2}\,\bar s_L\sigma_{\mu\nu} F^{\mu\nu}
    b_R \,, \nonumber\\
   O_8 &=& \frac{g_s m_b}{16\pi^2}\,\bar s_L\sigma_{\mu\nu}
    G_a^{\mu\nu} t_a b_R \,.
\end{eqnarray}
To an excellent approximation, the contributions of other operators
can be neglected. The renormalization scale $\mu_b$ in (\ref{Heff}) is
chosen of order $m_b$, so that all large logarithms reside in the
Wilson coefficients $C_i(\mu_b)$. For inclusive decays, the relevant
hadronic matrix elements of the local operators  $O_i$ can be
calculated using the heavy-quark expansion~\cite{Bigi}$^-$\cite{Adam}.
The complete theoretical prediction for the $B\to X_s\gamma$ decay
rate at next-to-leading order in $\alpha_s$ has been presented for the
first time by Chetyrkin et al.\ \cite{Chet}. It depends on a parameter
$\delta$ defined by the condition that the photon energy be above a
threshold given by $E_\gamma>(1-\delta) E_\gamma^{\rm max}$.  The
prediction for the $B\to X_s\gamma$ branching ratio is usually
obtained by normalizing the result for the corresponding decay rate to
that for the semileptonic rate, thereby eliminating a strong
dependence on the $b$-quark mass. We define
\begin{eqnarray}
   R_{\rm th}(\delta)
   &=& \frac{\Gamma(B\to X_s\gamma)\big|_{E_\gamma>(1-\delta)
             E_\gamma^{\rm max}}}
            {\Gamma(B\to X_c\,e\,\bar\nu)} \nonumber\\
   &=& \frac{6\alpha}{\pi f(z)}\,\left| \frac{V_{ts}^* V_{tb}}{V_{cb}}
    \right|^2 K_{\rm NLO}(\delta) \,,
\label{GNLO}
\end{eqnarray}
where $f(z)\approx 0.542-2.23(\sqrt z-0.29)$ is a phase-space factor
depending on the quark-mass ratio $z=(m_c/m_b)^2$. The fine-structure
constant $\alpha$ is renormalized at $q^2=0$, as is appropriate for
real-photon emission~\cite{CzMa}. The quantity $K_{\rm NLO}(\delta)$
contains the next-to-leading order corrections. In terms of the
theoretically calculable ratio $R_{\rm th}(\delta)$, the $B\to
X_s\gamma$ branching ratio is given by $\mbox{B}(B\to X_s\gamma)
=0.105 N_{\rm SL}\,R_{\rm th}(\delta)$, where $N_{\rm SL}
=\mbox{B}(B\to X_c\,e\,\bar\nu)/10.5\%$ is a normalization factor to
be determined from experiment. To good approximation $N_{\rm SL}=1$.
The current experimental situation of measurements of the semileptonic
branching ratio of $B$ mesons and their theoretical interpretation are
reviewed in Refs.~22,23.

In the calculation of the quantity $K_{\rm NLO}(\delta)$ we
consistently work to first order in the small parameters $\alpha_s$,
$1/m_Q^2$ and $\alpha/\alpha_s$, the latter ratio being related to the
leading-logarithmic QED corrections. The structure of the result is
\begin{equation}
   K_{\rm NLO}(\delta) = \sum_{ \stackrel{i,j=2,7,8}{i\le j} }
   k_{ij}(\delta,\mu_b)\,\mbox{Re}\!\left[ C_i(\mu_b)\, C_j^*(\mu_b)
   \right] \,,
\label{KNLO}
\end{equation}
where the Wilson coefficients $C_i(\mu_b)$ are expanded as
\begin{equation}
   C_i^{(0)}(\mu_b) + \frac{\alpha_s(\mu_b)}{4\pi}\,C_i^{(1)}(\mu_b) +
   \frac{\alpha}{\alpha_s(\mu_b)}\,C_i^{({\rm em})}(\mu_b) + \dots \,.
\label{Ciexp}
\end{equation}
The coefficients $C_i^{(k)}(\mu_b)$ are complicated functions of the
ratio $\eta=\alpha_s(m_W)/\alpha_s(\mu_b)$, which also depend on the
values $C_i(m_W)$ of the Wilson coefficients at the weak scale. In 
the Standard Model, these inital conditions are functions
of the mass ratio $x_t=(m_t/m_W)^2$. Whereas the leading-order
coefficients $C_i^{(0)}(\mu_b)$ are known since a long
time~\cite{Gr90}$^-$\cite{Poko}, the next-to-leading terms in
(\ref{Ciexp}), which must be kept for the coefficient $C_7(\mu_b)$,
have been calculated only recently. The expression for
$C_7^{(1)}(\mu_b)$ can be found in eq.~(21) of Ref.~16,
and the result for $C_7^{({\rm em})}(\mu_b)$ is given in eq.~(11) of
Ref.~18.

Explicit expressions for the functions $k_{ij}(\delta,\mu_b)$ in 
(\ref{KNLO}) can be found, e.g., in Ref.~18,
where we have 
corrected some mistakes in the formulae for real-gluon emission used 
by previous authors. (The corrected expressions are also given in the 
Erratum to Ref.~16.)
Bound-state corrections enter the formulae for the coefficients 
$k_{ij}$ at order $1/m_Q^2$ and are proportional to the hadronic
parameter~\cite{FaNe} $\lambda_2=\frac 14(m_{B^*}^2-m_B^2)\approx
0.12$\,GeV$^2$.  Most of them characterize the spin-dependent
interactions of the $b$ quark inside the $B$
meson~\cite{Bigi}$^-$\cite{Adam}. However, a  peculiar feature of
inclusive radiative decays is the appearance of a correction
proportional to $1/m_c^2$ in the coefficient $k_{27}$, which
represents a long-distance contribution arising from $(c\bar c)$
intermediate states~\cite{Volo}$^-$\cite{Buch}.

\subsection{Definition of the total branching ratio}

The theoretical prediction for the $B\to X_s\gamma$ branching ratio
diverges in the limit $\delta\to 1$ because of a logarithmic
soft-photon divergence of the $b\to sg\gamma$ subprocess, which would
be canceled by an infrared divergence of the $O(\alpha)$
corrections to the process $b\to s g$. We have argued that a
reasonable definition of the ``total'' branching ratio is to use an
extrapolation to $\delta=1$ starting from the region  $\delta\sim
0.5$--0.8, where the theoretical result exhibits a weak,  almost
linear dependence on the cutoff~\cite{us}. The extrapolated  value so
defined agrees, to a good approximation, with the result obtained by 
taking $\delta=0.9$, and hence we define the total branching ratio 
using this particular value of the cutoff.

The theoretical result is sensitive to the values of various input
parameters.   For the (one-loop) quark pole masses we take
$m_c/m_b=0.29\pm 0.02$, $m_b=(4.80\pm 0.15)$\,GeV, and $m_t=(175\pm
6)$\,GeV. The corresponding uncertainties in the  branching ratio are,
respectively, $\mbox{}^{+5.9}_{-5.0}\%$,  $\mp 1.0\%$, and $\pm
1.6\%$. We use the two-loop expression for the  running coupling
$\alpha_s(\mu)$ with the initial value $\alpha_s(m_Z)=0.118\pm 0.003$,
which induces an uncertainty of $\pm 2.7\%$. For the ratio of the CKM
parameters in (\ref{GNLO}) we take the value $|V_{ts}^*
V_{tb}|/|V_{cb}|=0.976\pm 0.010$ obtained from a global analysis of
the unitarity triangle~\cite{BurasWar}. This gives an uncertainty of
$\pm 2.1\%$. Finally, we include an uncertainty of $\pm 2.0\%$ to
account for next-to-leading electroweak radiative  corrections
\cite{CzMa}. The theoretical uncertainty arising from the variation
of the renormalization scale will be addressed below. We find an
uncertainty of $\pm 6.3\%$ from the variation of the scale $\mu_b$,
and of $\mbox{}^{+2.2}_{-1.5}\%$ from the variation of the scale
$\bar\mu_b$ entering the expression for the semileptonic decay rate in
the denominator in (\ref{GNLO}). Adding the different  errors in
quadrature gives a total uncertainty of
$\mbox{}^{+10}_{-\phantom{1}9}\%$ (adding them linearly would lead to
the more conservative estimate of $\mbox{}^{+24}_{-22}\%$). For the
total $B\to X_s\gamma$ branching ratio in the Standard Model we obtain
\begin{equation}
   \mbox{B}(B\to X_s\gamma)=(3.29\pm 0.33)\times 10^{-4} N_{\rm SL} \,,
\end{equation}
in good agreement with the experimental value in (\ref{BRexp}).

\subsection{Sensitivity to New Physics}

Possible New Physics contributions would enter  the theoretical
prediction for the $B\to X_s\gamma$ branching ratio through
non-standard  values of the Wilson coefficients of the dipole
operators $O_7$ and $O_8$ at the weak scale $m_W$.  To explore the
sensitivity to such effects, we normalize these coefficients to their
values in the Standard Model and introduce the ratios
$\xi_i=C_i(m_W)/C_i^{\rm SM}(m_W)$ with $i=7,8$.  In the presence of
New Physics,  the parameters $\xi_7$ and $\xi_8$ may take (even
complex) values different from 1. Similarly, New Physics may induce
dipole operators with opposite chirality to that of the Standard
Model, i.e.\ operators with right-handed light-quark fields. If we
denote by $C_7^R$ and $C_8^R$ the Wilson coefficients of these new
operators, expression (\ref{KNLO}) can be modified to include their
contributions by simply replacing $C_i C_j^*\to C_i C_j^* + C_i^R
C_j^{R*}$ everywhere.  We thus define two additional parameters
$\xi_i^R=C_i^R(m_W)/C_i^{\rm SM}(m_W)$ with $i=7,8$, which vanish in
the Standard Model. Since the dipole operators only contribute to rare
flavour-changing neutral current processes, there are at present
rather weak constraints on the values of these
parameters~\cite{JoAnne}. On the other hand, we assume that the
coefficient $C_2$ of the current--current operator $O_2$ takes its
standard value, and that there is no similar operator containing
right-handed quark fields.  With these definitions, the $B\to
X_s\gamma$ branching ratio can be decomposed as~\cite{us}
\begin{eqnarray}
   &&\hspace{-0.4cm} \frac{1}{N_{\rm SL}}\,{\rm B}(B\to X_s\gamma)
    \big|_{E_\gamma>(1-\delta)E_\gamma^{\rm max}} \nonumber\\ &=&
    B_{22}(\delta) + \!B_{77}(\delta)\!\left( |\xi_7|^2 + |\xi_7^R|^2
    \right) + \!B_{88}(\delta)\!\left( |\xi_8|^2 + |\xi_8^R|^2 \right)
    \nonumber\\ &&\mbox{}+ B_{27}(\delta)\,\mbox{Re}(\xi_7) +
    B_{28}(\delta)\,\mbox{Re}(\xi_8) \nonumber\\ &&\mbox{}+
    B_{78}(\delta)\!\left[ \mbox{Re}(\xi_7\xi_8^*) +
    \mbox{Re}(\xi_7^R\xi_8^{R*}) \right] \,.
\label{xidecomp}
\end{eqnarray}
In Table~\ref{tab:Bij}, the values of the components $B_{ij}(\delta)$
obtained with $\delta=0.9$ are shown for different choices of the
renormalization scale. Assuming $N_{\rm SL}=1$, the Standard Model
result for the total branching ratio is given by $\mbox{B}_{\rm
SM}=\sum_{ij} B_{ij}$. The most important contributions are $B_{27}$
and $B_{22}$ followed by $B_{77}$. The smallness of the remaining
terms shows that there is little sensitivity to the coefficient
$C_8(m_W)$ of the chromo-magnetic dipole operator. Once the parameters
$\xi_i$ and $\xi_i^R$ are calculated in a given New Physics scenario,
the result for the $B\to X_s\gamma$ branching ratio can be derived
using the numbers shown in the table. For the remainder of this talk,
however, we assume the validity of the Standard Model.

\begin{table}
\caption{Values of the coefficients $B_{ij}$ (for $\delta=0.9$) in
units of $10^{-4}$, for different choices of $\mu_b$. The coefficient
$B_{88}=0.015$ in all three cases}
\vspace{0.2cm}
\begin{center}
\begin{tabular}{|c|cccccc|}
\hline   
$\mu_b$ & $B_{27}$ & $B_{22}$ & $B_{77}$ & $B_{28}$ & $B_{78}$ &
 $\!\!\!\!\sum_{ij} B_{ij}\!\!$ \\   
\hline\hline   
$m_b/2$ & 1.265 & 1.321 & 0.335 & 0.179 & 0.074 & 3.188 \\   
$m_b$ & 1.395 & 1.258 & 0.382 & 0.161 & 0.083 & 3.293 \\   
$2 m_b$ & 1.517 & 1.023 & 0.428 & 0.132 & 0.092 & 3.206 \\ 
\hline
\end{tabular}
\end{center}
\label{tab:Bij}
\end{table}

\subsection{Perturbative uncertainties}

The components $B_{ij}(\delta)$ in (\ref{xidecomp}) are  formally
independent of the renormalization scale $\mu_b$. Their residual scale
dependence results only from the truncation of perturbation theory at
next-to-leading order and is conventionally taken as an estimate of
higher-order corrections. Typically, the different components vary by
amounts of order 10--20\% as $\mu_b$ varies between $m_b/2$ and
$2m_b$. The good stability is a result of the cancelation of the scale
dependence between Wilson coefficients and matrix elements achieved by
a next-to-leading order calculation.

Previous authors~\cite{Buras,Chet,Ciuc,Borzu} who estimated the
$\mu_b$ dependence of the total $B\to X_s\gamma$  branching ratio in
the Standard Model found a more striking improvement over the
leading-order result, namely a variation of only
$\mbox{}^{+0.1}_{-3.2}\%$ as compared with $\mbox{}^{+27}_{-20}\%$ at
leading order. However, the apparent excellent stability observed at
next-to-leading order is largely due to an accidental
cancelation between different contributions to the branching ratio.  A
look at Table~\ref{tab:Bij} shows that the residual scale dependence
of the individual contributions $B_{ij}$ is much larger than that of
their sum, which determines the total branching ratio in the Standard
Model. Note, in particular, the almost perfect cancelation of the
scale dependence between $B_{27}$ and $B_{22}$, which is
accidental since the magnitude of $B_{27}$ depends on the
top-quark mass, whereas $B_{22}$ is independent of $m_t$. In such
a situation, the apparent weak scale dependence of the sum of all
contributions is not a good measure of higher-order corrections.
Indeed, higher-order corrections must stabilize the different
components $B_{ij}$  individually, not only their sum. The variation
of the individual components as a function of $\mu_b$ thus provides a
more conservative estimate of the truncation error than does the
variation of the total branching ratio. For each component, we
estimate the truncation error by taking one half of the maximum
variation obtained by varying $\mu_b$ between $m_b/2$ and $2m_b$. The
truncation error of the sum is then obtained by adding the individual
errors in quadrature. In this way, we find a total truncation error of
$\pm 6.3\%$, which is more than a factor of 2 larger than the
estimates obtained by previous authors. An even larger truncation
error could be  justified given that the choice of the range of
variation of $\mu_b$  is ad hoc, and that the scale dependence of the
various components is not symmetric around the point $\mu_b=m_b$.

\section{Partially integrated branching ratio and photon-energy 
spectrum}

Whereas the explicit power corrections to the functions $k_{ij}$ are
small, an important nonperturbative effect not  included so far is the
motion of the $b$ quark  inside the $B$ meson caused by its soft
interactions with the light constituents. It leads to a modification
of the photon-energy spectrum, which must be taken into account  if a
realistic cutoff is imposed~\cite{us,Alex98}. This so-called ``Fermi
motion'' can be included in the heavy-quark expansion by resumming an
infinite set of leading-twist contributions into a shape function
$F(k_+)$, which governs the light-cone momentum distribution of the
heavy quark inside the $B$ meson~\cite{shape}$^-$\cite{shape2}.  The
physical decay distributions are obtained from a convolution of
parton-model spectra with this function.  In the process, phase-space
boundaries defined by parton kinematics are transformed into the
proper physical boundaries defined by hadron kinematics.

The shape function is a universal characteristic of the $B$ meson
governing the inclusive decay spectra in processes with massless
partons in the final state, such as $B\to X_s\gamma$ and $B\to
X_u\,\ell\,\nu$.  However, this function does not describe in an
accurate way the distributions in decays into massive partons such
as~\cite{Russi,shape2} $B\to X_c\,\ell\,\nu$. Unfortunately,
therefore, the shape function cannot be determined using the lepton
spectrum in semileptonic decays, for which high-precision data
exist. On the other hand, there is some useful theoretical information
on the moments $A_n=\langle k_+^n\rangle$ of the function $F(k_+)$,
which are related to the forward matrix elements of local
operators~\cite{shape}. In particular, $A_1=0$  vanishes by the
equations of motion (this condition defines the heavy-quark mass), 
and $A_2=\frac 13\mu_\pi^2$ is related to the kinetic energy of the 
heavy quark inside the $B$ meson. For our purposes it is sufficient 
to adopt the simple form $F(k_+) = N\,(1-x)^a e^{(1+a)x}$ with $x =
k_+/\bar\Lambda\le 1$,  where $\bar\Lambda=m_B-m_b$. This ansatz is
such that $A_1=0$ by construction. The parameter $a$ can be related to
the second moment, yielding $\mu_\pi^2=3\bar\Lambda^2/(1+a)$. Thus,
the $b$-quark mass (or $\bar\Lambda$) and  the quantity $\mu_\pi^2$
(or $a$) are the two parameters of our  function. We take
$m_b=4.8$\,GeV and $\mu_\pi^2=0.3$\,GeV$^2$ as reference values, in
which case $a\approx 1.29$.

Let us denote by $B_{ij}^{\rm p}(\delta_{\rm p})$ the various
components in (\ref{xidecomp}) calculated in the parton model, where
the cutoff $\delta_{\rm p}$ is defined by the condition that
$E_\gamma\ge\frac 12(1-\delta_{\rm p}) m_b$. Then the corresponding
physical quantities $B_{ij}(\delta)$ with $\delta$ defined such that
$E_\gamma\ge\frac 12(1-\delta) m_B$ are given by~\cite{us,Alex98}
\begin{equation}
   B_{ij}(\delta) = \!\int\limits_{m_B(1-\delta)-m_b}^{m_B-m_b}\!\!\!
   \mbox{d}k_+\,F(k_+)\,B_{ij}^{\rm p}\!\left( 1 -
   \frac{m_B(1-\delta)}{m_b+k_+} \right) \,.
\label{recipe}
\end{equation}
This relation is such that $B_{ij}(1)=B_{ij}^{\rm p}(1)$, implying
that the total branching ratio is not affected by Fermi motion.  The
effect is, however, important for realistic values of the cutoff
$\delta$.

\begin{figure}
\epsfxsize=8.7cm   
\centerline{\epsffile{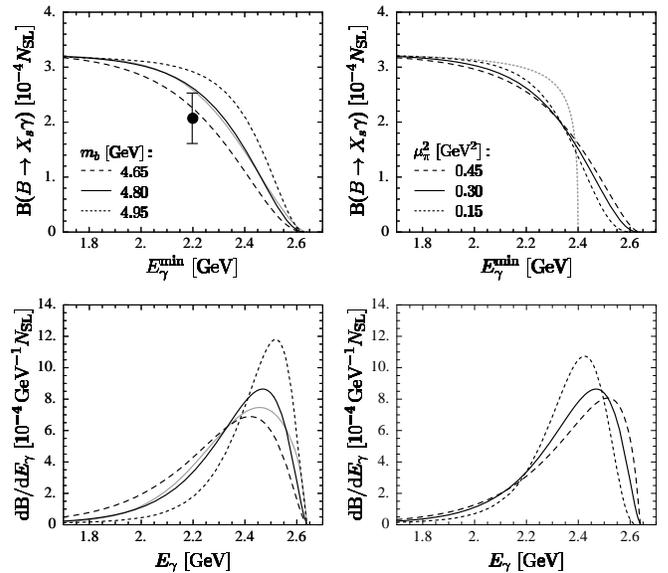}}
\caption{Partially integrated $B\to X_s\gamma$ branching ratio (upper
plots) and photon-energy spectrum (lower plots) for various choices of
shape-function parameters}
\label{fig:Fermi_motion}
\end{figure}

As an illustration of the sensitivity of our results to the parameters
of the shape function, the upper plots in
Figure~\ref{fig:Fermi_motion} show the predictions for the  partially
integrated $B\to X_s\gamma$  branching ratio as a function of the
energy cutoff $E_\gamma^{\rm min}=\frac 12(1-\delta) m_B$. In the
left-hand plot we vary $m_b$ keeping the ratio
$\mu_\pi^2/\bar\Lambda^2$ fixed.  The gray line shows the result
obtained using the same parameters as for the solid line but a
different functional form given by  $F(k_+)=N\,(1-x)^a e^{-b(1-x)^2}$.
For comparison, we show the data point $\mbox{B}(B\to
X_s\gamma)=(2.04\pm 0.47)\times 10^{-4}$ obtained in the original CLEO
analysis~\cite{CLEO} with a cutoff at 2.2\,GeV.  In the right-hand
plot, we keep $m_b=4.8$\,GeV fixed and compare the  parton-model
result (gray dotted curve) with the results corrected for Fermi motion
using different values for the parameter $\mu_\pi^2$.  This plot
illustrates how Fermi motion fills the gap between the parton-model
endpoint at $m_b/2$ and the physical endpoint at $m_B/2$. (The  true
endpoint is actually located at $[m_B^2-(m_K+m_\pi)^2]/2 m_B\approx
2.60$\,GeV, i.e.\ slightly below $m_B/2\approx 2.64$\,GeV.) Comparing
the two plots, it is evident that the uncertainty due to the value of
the $b$-quark mass is the dominant one. Variations of the parameter
$\mu_\pi^2$ have a much smaller effect on the partially integrated
branching ratio, and also the sensitivity to the functional form
adopted for the shape function turns out to be small. This behaviour
is a consequence of global quark--hadron duality, which ensures that
even partially integrated quantities are rather insensitive to
bound-state effects. The strong remaining dependence on the $b$-quark
mass is simply due to the transformation by Fermi motion of
phase-space boundaries from parton to hadron kinematics.

\begin{table}
\caption{Partially integrated $B\to X_s\gamma$ branching ratio for
different values of the cutoff on the photon energy}
\vspace{0.2cm}
\begin{center}
\begin{tabular}{|c|c|}
\hline  
$E_{\gamma,{\rm min}}^{\rm lab}$ & $\mbox{B}(B\to X_s\gamma)~
[10^{-4} N_{\rm SL}]$ \\
\hline\hline  
2.2\,GeV & $2.56\pm 0.26\,\mbox{}^{+0.31}_{-0.36}$ \\  
2.1\,GeV & $2.85\pm 0.29\,\mbox{}^{+0.18}_{-0.27}$ \\  
2.0\,GeV & $3.01\pm 0.30\,\mbox{}^{+0.09}_{-0.18}$ \\
\hline
\end{tabular}
\end{center}
\label{tab:Brs}
\end{table}

Taking the three curves in the left-hand upper plot for a
representative range of parameters and applying a small correction for
the boost from the $B$ rest frame to the laboratory as appropriate for
the CLEO analysis, we show in Table~\ref{tab:Brs} the predictions  for
the partially integrated branching ratio for three different values of
the cutoff on the photon energy. The first error refers to the
dependence on the various input parameters discussed previously, and
the second one accounts for the  uncertainty associated with the
description of Fermi  motion. In general,  this second error can be
reduced in two ways. The first possibility is to lower the cutoff on
the photon energy. A first step in this direction has already been
taken in the new CLEO analysis reported at this
Conference~\cite{CLEOnew},  in which the cutoff has been lowered from
2.2 to 2.1\,GeV. If a value as low as 2\,GeV could be achieved, the
theoretical predictions would become rather  insensitive to the
parameters of the shape function. To what extent this will be possible
in future experiments will depend on their capability to reject the
background of photons from other decays. The second possibility is
that future high-precision measurements of the photon spectrum will
make it possible to adjust the parameters of the shape function from a
fit to the data. For the purpose of  illustration, the photon spectra
corresponding to the  various parameter sets are shown in the lower
plots in  Figure~\ref{fig:Fermi_motion}. Such  a determination of the
shape-function parameters from $B\to X_s\gamma$ decays would not only
help to reduce the theoretical uncertainty in the determination of the
total branching ratio, but would also enable us to predict
the lepton spectrum in $B\to X_u\,\ell\,\nu$ in a model-independent
way~\cite{shape}. This may help to reduce the theoretical uncertainty
in the value of $|V_{ub}|$. A detailed analysis of the photon
spectrum will therefore be an important aspect in future analyses of
inclusive radiative $B$ decays.

\section{Conclusions}

The inclusive radiative decays $B\to X_s\gamma$ play a key role in
testing the Standard Model and probing the structure of possible New
Physics.  A reliable theoretical calculation of their branching ratio
can be performed using the operator product expansion  for inclusive
decays of heavy hadrons combined with the twist expansion for the
description of decay distributions near phase-space boundaries.  To
leading order in $1/m_b$ the decay rate agrees with the parton-model
rate for the underlying quark decay $b\to X_s\gamma$.  With the
completion of the next-to-leading order calculation of the Wilson
coefficients and matrix elements  of the operators in the effective
weak Hamiltonian,  the perturbative uncertainties in the calculation
of this process  have been reduced to a level of about 10\%.
Bound-state corrections to the total decay rate are suppressed by
powers of $1/m_b$ and can be controlled in a systematic way.

A more important effect is the Fermi motion of the heavy quark inside
the meson, which is responsible for  the characteristic shape of the
photon-energy spectrum in $B\to X_s\gamma$ decays.  It leads to the
main theoretical uncertainty in the calculation of the branching ratio
if a restriction to the high-energy part of the photon spectrum is
imposed. Fermi motion is naturally incorporated in the heavy-quark
expansion by resumming an infinite set of leading-twist operators into
a non-perturbative shape function. The main theoretical uncertainty in
this description lies in the value of the $b$-quark mass. Other
features associated with the detailed functional form of the shape
function play a minor role. The value of $m_b$ and other
shape-function  parameters could, in principle, be extracted from a
precise measurement  of the photon-energy spectrum, but also gross
features of this spectrum  such as the average photon energy would
provide valuable  information~\cite{us}. For completeness, we note
that besides the  photon-energy spectrum also the invariant hadronic
mass distribution in  radiative $B$ decays can be studied. 
Investigating the pattern of individual hadron resonances contributing 
to the spectrum, one can motivate a simple description of the hadronic 
mass spectrum with only a  single parameter, the $B\to K^*\gamma$ 
branching ratio, to be determined from experiment~\cite{us}.

Possible New Physics contributions would enter the theoretical
predictions  for the $B\to X_s\gamma$ branching ratio and photon
spectrum  through the values of parameters $\xi_i$ and $\xi_i^R$, which 
are defined in terms of values of Wilson coefficients at the
scale $m_W$. This formalism allows one to account for non-standard
contributions to the magnetic and chromo-magnetic dipole operators, as
well as operators with right-handed light-quark fields.  Quite
generally, New Physics would not affect the shape of the photon
spectrum~\cite{us} but could change the total branching ratio by a
considerable amount. This implies that the analysis of the
photon-energy and hadronic mass spectra, which is crucial for the
experimental determination of the total branching ratio, can be
performed without assuming the correctness of the Standard Model. On
the other hand, the total branching ratio will provide a powerful
constraint on the structure of New Physics beyond the Standard Model
as experimental data become more precise.

\section*{Acknowledgments}

The work reported here has been done in a most pleasant collaboration
with Alex Kagan, which is gratefully acknowledged.

\end{document}